\documentstyle[aps,preprint]{revtex}
\begin{document}
\draft
\title{Spin content of the nucleon in a valence and sea quark mixing model}

\author{Di Qing, Xiang-Song Chen and Fan Wang}
\address{Department of Physics and Center for Theoretical Physics, 
Nanjing University, Nanjing, 210093, P.R. China}

\date{\today}
\maketitle

\begin{abstract}
A dynamical valence and sea quark mixing model is shown to fit the baryon ground 
state
properties as well as the spin content of the nucleon. The relativistic
correction and the $q^3{\leftrightarrow}q^3q\bar{q}$ transition terms 
induced by the quark
axial vector current $\bar{\psi}\vec{\gamma}\gamma^5\psi$ in this model
space is responsible for the quark spin
reduction. 
\end{abstract}


The naive valence quark model after incorporating QCD effective one gluon
exchange and phenomenological confinement interactions is quite successful in
explaining hadron properties [1] and is encouraging in describing 
hadron interactions [2]. Therefore it  seems to be a good model of 
hadron internal structure especially for the nucleon. The EMC measurement [3]
shows only a small amount of the nucleon spin is carried by the
quark spin. This surprising result challenges our understanding of 
nucleon structure and has stimulated a new round of nucleon structure
studies. The vast literature can be found from the invited talks given at
recent conferences [4]. We only mention a few which are relevant to the
present discussion. Jaffe and Lipkin [5] proposed a toy model with
$q^3$ and $q^3q\bar{q}$ mixing to accommodate the EMC result. Hwang, Speth
and Brown [6] used the generalized Sullivan processes with 
phenomenological meson-baryon coupling vertices to explain the spin-flavor
structure of the nucleon. Cheng and Li [7] used the chiral quark
model to remedy the failures of the naive quark model. Ma and Brodsky [8] 
emphasized the relativistic reduction of the quark spin contribution due to 
the Melosh
rotation and included a small amount of the intrinsic sea quark component
caused by the energetically-favored meson-baryon fluctuations to explain the
violation of Ellis-Jaffe sum rule and Gottfried sum rule. Close [9]
reiterated that the polarization asymmetry in the valence region confirms the
naive valence quark model predictions and one should focus on the sea quark
polarization especially the small $x$ behaviour.

  There have been various suggestions to include the gluon spin and the quark
and gluon orbital angular momentum contributions in the nucleon spin. However as clarified
by Ji [10] and ourselves [11], in the usual decomposition of the nucleon
spin,
\begin{eqnarray}
       && {1\over 2} = {1\over 2}\Delta \Sigma + \Delta G+L_q+L_G,\nonumber \\
       && \Delta \Sigma = \langle p+\left| \int d^3x
\bar{\psi}\gamma^3\gamma^5\psi \right| p+\rangle \nonumber, \\
       && \Delta G = \langle p+\left| \int d^3x (E^1A^2-E^2A^1)
\right| p+\rangle, \nonumber \\
       && L_q = \langle p+\left|\int d^3x {1\over i}\psi^{\dag}(x^1\partial^2 -
x^2\partial^1)\psi\right| p+\rangle, \nonumber \\
       && L_G = \langle p+\left|\int d^3x E^i(x^1\partial^2 -
x^2\partial^1)A^i\right| p+\rangle ,
\end{eqnarray}
the terms, except the $\Delta \Sigma$ term, are neither
separately gauge invariant nor Lorentz invariant. The gauge invariance is
obvious, and the Lorentz invariance can be expressed as
\begin{equation}
  \Delta \Sigma s^\mu = \langle ps\left|\int
d^3x\bar{\psi}\gamma^\mu\gamma^5\psi \right|ps\rangle .
\end{equation}

   The quark and gluon contribution to the nucleon spin can be decomposed in
the gauge invariant formalism as
\begin{equation}
  \vec{J} = \int d^3x\psi^{\dag}{1\over 2}\vec{\Sigma}\psi+\int d^3x
\psi^{\dag}\vec{x}\times{1\over i}\vec{D}\psi+\int d^3x \vec{x}
\times(\vec{E}\times\vec{B}).
\end{equation}
Here $\vec{D}$ is the covariant derivative, but
$\vec{r}\times{1\over i}\vec{D}$ does not obey the angular momentum
commutation relation.

  The third term is the gluon contribution, including both the gluon spin and
orbital angular momentum, and it is impossible to decompose this term
into individually gauge invariant gluon spin and orbital angular momentum 
parts.

  Due to these uncertainties we will concentrate our discussion on the 
contribution from
the quark axial vector current operator  
   $\int d^3x\bar{\psi}\vec{\gamma}\gamma^5\psi $.
In the parton model manifested at infinite momentum frame
\begin{equation}
\Delta\Sigma = \int dx\left(q^\uparrow\left(x\right)-q^\downarrow\left(x\right)\right),
\end{equation}
where $q^{\uparrow,\downarrow}\left(x\right)$ is the probability of finding a
quark or antiquark with fraction x of the proton longitudinal momentum 
and polarization parallel or antiparallel to the proton spin.

It is a quite intuitive impression from eq.(4) that the counterpart of 
$\Delta\Sigma$ in the nonrelativistic constituent quark model is:
\begin{equation}
\Delta\Sigma^{NR} = \int d^3p\left(q^\uparrow\left(\vec p\right)-q^\downarrow\left(\vec p\right)\right),
\end{equation}
where $q^{\uparrow,\downarrow}\left(\vec p\right)$ is the probability of finding a
quark or antiquark  of momentum $\vec p$ and 
polarization parallel or antiparallel to the proton spin in the nonrelativistic
constituent quark model manifested at the proton rest frame.

This mis-identifying eq.(5) is the root of the confusion related to the nucleon
spin structure. We will show that eq.(5) is true only for a static valence 
$\left(q^3\right)$ quark model. For any realistic QCD inspired quark model
eq.(5) is not true.

In the following discussion we will still use the term "quark spin contribution 
to the nucleon spin". In fact we are always talking about the matrix element 
of the quark axial vector current operator, which is the quantity measured in
the deep inelastic scattering.
To evaluate the axial vector current operator (2) in a
nonrelativistic constituent quark model, we assume the quark field operator 
$\psi$ can be directly
related to the constituent quark degree of freedom. This is a usual
assumption in such a model calculation, but needs to be studied further [12].
The next step is simple but seems to be missed in a few model calculations
[13]. The nonrelativistic reduction of the current operator includes not only
the Pauli spin operator but also a relativistic correction term
\begin{equation}
   \int d^3x\bar{\psi}\vec{\gamma}\gamma^5\psi = \sum_{s^\prime
s}\int d^3p\chi^{\dag}_{s^\prime}(\vec{\sigma}+{\vec{\sigma}\cdot\vec{p}\over
2E(E+m)}[\vec{\sigma},\vec{\sigma}\cdot\vec{p}])\chi_sa^+_{ps^\prime}a_{ps}.
\end{equation}

  This kind of relativistic reduction was discussed earlier in a pure
phenomenological manner [14]. Applying this to the Isgur model [1],
we have
\begin{equation}
   \Delta \Sigma = (1-{1 \over 3m^2b^2})\sim 0.68.
\end{equation}
Therefore the matrix element of the axial vector current operator, which 
is called the quark spin contribution in the literature, in a 
nonrelativistic model is not
$\Delta \Sigma = 1$ but around 0.70. This is due to quark Fermi motion in a
confined region $b$. Only in the static $SU_6^{\sigma f}$ model, i.e.,the 
case in which all the
internal quark momenta $\vec{p}=0$, has one $\Delta \Sigma=1$. This
result is similar to that of Ma and Brodsky [8] based on Melosh rotation
and a light cone formalism.
  
  The world average value of $\Delta \Sigma$ is [15]
\begin{eqnarray}
\Delta \Sigma(Q^2\sim3GeV^2) &=& \Delta u+\Delta d+\Delta s \nonumber \\
  &=& 0.81(\pm0.01)-0.44(\pm0.01)-
   0.10(\pm0.01)=0.27(\pm0.04).
\end{eqnarray}
A possible contribution to the remaining difference ($\Delta\Sigma=0.68-0.27$) 
is the intrinsic sea quark component of the nucleon [5-9].
In the following we use a dynamical valence and sea quark mixing 
model [16]
to study this problem.

  In order to keep the successful part of the naive valence quark model, we
assume a model Hamiltonian quite similar to that of the Isgur model [1].
However a new ingredient, the sea quark excitation interaction, is
introduced in order to mix the $q^3q\bar{q}$ configuration with the $q^3$ 
valence
part. Such a Hamiltonian should be written in a second quantized 
formalism, but we still use first quantization with an understanding
that the one and two body operators include different particle numbers in
different sub-Hilbert spaces.
\begin{eqnarray}
     && H=\sum_i(m_i+{p^2_i\over 2m_i})+\sum_{i<j}(V_{ij}^c+V^G_{ij})+
\sum_{i<j}(V_{i,i^\prime j^\prime j}+V^{\dag}_{i,i^\prime j^\prime j})\nonumber,  \\
     && V^c_{ij}=-a_c\vec{\lambda}_i\cdot\vec{\lambda}_jr^2_{ij}\nonumber, \\
     && V_{ij}^{Gs}=\alpha_s{\vec{\lambda}_i\cdot\vec{\lambda}_j\over 4}
({1\over r_{ij}}-{\pi\over 2}({1\over m^2_i}+{1\over m^2_j}+{4\vec{\sigma}_i
\cdot\vec{\sigma}_j\over 3m_im_j})\delta(\vec{r}_{ij})+\cdot\cdot\cdot)
\nonumber, \\
     && V_{ij}^{Ga}=\alpha_s({\vec{\lambda}_i\cdot\vec{\lambda}_j^*\over 2})^2
({1\over 3}+{\vec{f}_i\cdot\vec{f}_j^*\over 2})({\vec{\sigma}_i\cdot
\vec{\sigma}_j\over 2})^2{2\over 3}{1\over (m_i+m_j)^2}\delta(\vec{r}_{ij})
\nonumber, \\
     && V_{i,i^\prime j^\prime j}=i\alpha_s{\vec{\lambda}_i\cdot\vec{\lambda}_j
\over 4}{1\over 2r_{ij}}((({1\over m_i}+{1\over m_j})\vec{\sigma}_j+
{i\vec{\sigma_j}\times\vec{\sigma_i}\over m_i})\cdot{\vec{r}_{ij}\over
r^2_{ij}} -{2\vec{\sigma}_j\cdot\vec{\nabla}_i\over m_i}), 
\end{eqnarray}
where $\vec{\lambda}_i(\vec{f}_i)$ are the $SU^c_3(SU^f_3)$ Gellmann
operators, the $V^{Gs}_{ij}$, $V^{Ga}_{ij}$ and $V_{i,i^\prime j^\prime j}$
correspond to the following diagrams of Fig.1 respectively, the other 
symbols have their usual meaning.

  \fbox{fig.1 goes here.}

  Following the chiral quark model [17], the model Hilbert space is
truncated to a subspace which includes all possible combinations of 
color singlet s-wave
$q^3$ baryon states and $^1S_0$ $q\bar{q}$ pseudo scalar meson states
compatible with the quantum number of a baryon.
The color, spin, flavor wave functions of the $q^3$ baryon core and the 
$q\bar{q}$ meson are the usual $SU^c_3\times SU^{\sigma f}_6$ ones. The
internal orbital wave functions of $q^3$ and $q\bar{q}$ are assumed to be
a Gaussian with a common size parameter $b$. The relative motion between
$q^3$ baryon core and $q\bar{q}$ meson is assumed to be a
p-wave to meet the
positive parity requirement of ground state baryons. For simplicity, it is
assumed to be a p-wave Gaussian with the same $b$ as that of the internal
part. Essentially we use a shell model approximation but the wave function 
of the center of mass is eliminated.

  The model parameters, $u,d$ quark mass $m$, s quark mass $m_s$, 
quark gluon coupling constant $\alpha_s$, $q^3$ quark core baryon size 
$b$, and confinement
strength $a_c$, are fixed by an overall fit to the ground state octet and
decuplet baryon masses and the magnetic moments of the octet. The root mean
square charge radius of proton is also fitted. A relativistic correction
term (to the order of ${p^2\over m^2}$)is included in the calculation of the
nucleon charge radius.

  \fbox{Table I goes here.}

  \fbox{Table II goes here.}

  Table I shows the wave function of the proton. The entry is the amplitude
of the individual component. It is an example of our
model wave functions of ground state baryons.

  Table II summarize our model predictions and the model parameters. These
results show that it is possible to have a valence and sea quark mixing model
which can describe, with the commonly accepted quark model parameters, 
the ground state octet and decuplet baryon properties 
as good as
the successful naive valence quark
model.  Furthermore, the proton charge radius is reproduced as well. The
first excited states are higher than 2 GeV. This is consistent with the fact
that there is no pentaquark states observed below 2 GeV.

  \fbox{Table III goes here.}

The spin structure of the proton is listed in table III, where the matrix
element of the axial vector current operator (2) in a spin up proton state
is decomposed into particle number conserved components $q^3 \leftrightarrow
q^3, q^4\bar{q} \leftrightarrow q^4\bar{q}$ and particle number 
nonconserverd components
$q^3 \leftrightarrow q^4\bar{q}$.  
The relativistic correction (6) has been taken into
account in the calculation of the $q^3 \leftrightarrow q^3$ matrix element. 
After
antisymmetrization, it is impossible to separate the $u,d$ valence and sea
quark contribution of $q^3q\bar{q}$ components. Moreover 
in addition to the particle number conserved term (6),
due to 
mixing of $q^3$ and $q^4\bar{q}$ components the axial vector operator 
has a particle number nonconserved term between $q^3$ and $q^4\bar{q}$ components,
\begin{equation}
  \int d^3x\bar{\psi}\vec{\gamma}\gamma^5\psi=\sum_{s,s^\prime} \int d^3p
\chi^{\dag}_{s^\prime} i {\vec{\sigma}\times\vec{p}\over E}\chi_s a^+_{ps^\prime}
b^+_{-ps},
\end{equation}
where $b^+_{-ps}$ is antiquark creation
operator. This particle number nonconserved term (and its Hermitian 
conjugate) gives rise an
additional contribution to the nucleon spin. It is this transition term which
contributes negative $\Delta q$, which in turn reduces the $\Delta\Sigma$ 
of proton
further. Physically, this transition term is similar to the generalized 
Sullivan processes which has been discussed in [6]. Adding these three 
contributions together, we obtain a spin
distribution $\Delta u,\Delta d$ and $\Delta s$ quite close to the world
average result.
  
Our conclusion is that a nonrelativistic quark model with small amount of
$q^3q\bar{q}$ component mixing is able to explain the $\Delta\Sigma(Q^2\sim
3GeV^2)\sim 0.27$ measured in the deep inelastic scattering and at the
same time keep a good fit to the baryon properties. 
The
key point is to distinguish the quark spin sum which is 1 for a pure
valence quark model from the matrix element of the quark axial vector 
current operator which is measured
in the deep inelastic scattering. As for the nucleon spin, i.e., the total 
angular momentum of the nucleon, we should point out that it is still 
$1\over 2$ in our scheme. Because the content of quark orbital angular momentum in QCD is also different from that in nonrelativistic quark model, and if we 
make the nonreltivistic reduction of it, we will get relativistic correction 
terms as well. Simply speaking, these correction terms come from the small 
component 
of Dirac spinors. Furthermore, they are exactly the same but with opposite 
sign as the correction terms from the quark axial vector current, 
therefore guarantee the 
nucleon spin to be $1\over 2$.

  It should be mentioned that we have not adjusted the parameters very
carefully for getting a perfect fit,
since our aim is to show that the nucleon spin content measured in the 
deep inelastic scattering is understandable in a nonrelativistic quark 
model. Our model itself is a very rough one.  Firstly, the $q^3$ and 
$q^3q\bar{q}$ mixing
interaction is derived by an effective one gluon exchange while the real
interaction is quite likely to be nonperturbative. Secondly, in our model
the pseudo scalar meson is approximated as a pure $q\bar{q}$ state and 
 only the pseudo scalar meson is included in our truncated space,
which is rather artificial. If the space is enlarged to 
include vector meson, we found
that $N\omega, N\rho, \Delta\rho, \Lambda K^*$ components are mixed as
strongly as the pseudo scalar ones
and the fit is not
better but even worse. Another point worth mentioning is that 
the shell model approximation of the orbital wave
function is questionable. In fact it should be a meson baryon continuum. The
relativistic correction is also questionable quantitatively, since in our
model the ${p\over m}$ is not small. Certainly much work should be done
in the future.

This work is supported by the NSF (19675018), SEDC and SSTC of China.

\pagebreak
\begin{center}
FIGURE CAPTION
\vspace{3cm}

Fig.1 quark interaction diagrams
\end{center}
\begin{table}
\begin{center}
\caption{proton model wave function}
\begin{tabular}{|c|c|c|c|c|c|c|c|}
%\begin{tabular}{cccccccc}
$q^3$&$N\eta$&$N\pi$&$\Delta\pi$&$N\eta^\prime$&$\Lambda K$&$\Sigma K$
&$\Sigma^* K$ \\ \hline
$-$0.923&0.044&0.232&$-$0.252&0.065&0.109&$-$0.036&$-$0.106 \\ 
\end{tabular}
\end{center}
\end{table}

\begin{table}
\begin{center}
\caption{masses and magnetic moments of the baryon
octect and decuplet.
$m=330(MeV),m_s=564(MeV),b=0.61(fm),\alpha_s=1.46,a_c=48.2(MeVfm^{-2})$}
\begin{scriptsize}
\begin{tabular}{|c|c|c|c|c|c|c|c|c|c|c|c|c|}
%\begin{tabular}{ccccccccccccc}
    &   &p  &n   &$\Lambda$&$\Sigma^+$&$\Sigma^-$&$\Xi^0$&$\Xi^-$&
 $\Delta$&$\Sigma^*$&$\Xi^*$&$\Omega$ \\ \hline
    &M(Mev)&\multicolumn{2}{c|}{939}&1116&\multicolumn{2}{c|}{1193}
 &\multicolumn{2}{c|}{1346}&1232&1370&1523&1659 \\ \cline{2-13}
Theor.&E1(MeV)&\multicolumn{2}{c|}{2203}&2323&\multicolumn{2}{c|}{2306}
&\multicolumn{2}{c|}{2409}&2288&2306&2450&2638 \\ \cline{2-13}
    &$\mu(\mu_N)$&2.780&$-$1.818&$-$0.522&2.652&$-$1.072&$-$1.300&$-$0.412
& & & & \\
\cline{2-13}
    &$\sqrt{\langle r^2 \rangle}(fm)$&0.802&0.124& & & & & & & & & \\ \hline
    &M(MeV)&\multicolumn{2}{c|}{939}&1116&\multicolumn{2}{c|}{1189}
&\multicolumn{2}{c|}{1315}&1232&1385&1530&1672 \\ \cline{2-13}
Exp.&$\mu(\mu_N)$&2.793&$-$1.913&$-$0.613&2.458&$-$1.160&$-$1.250&$-$0.651
& & & & \\
\cline{2-13}
    &$\sqrt{\langle r^2 \rangle}(fm)$&0.836&0.34& & & & & & & & & \\ 
\end{tabular}
\end{scriptsize}
\end{center}
\end{table}

\begin{table}
\begin{center}
\caption{The spin content of proton}
\begin{tabular}{|c|c|c|c|c|c|}
%\begin{tabular}{cccccc}
  &$q^3$&$q^3-q^4\bar{q}$&$q^4\bar{q}-q^4\bar{q}$&sum&exp. \\ \hline
$\Delta u$&0.773&$-$0.125&0.143&0.791&0.81 \\ \hline
$\Delta d$&$-$0.193&$-$0.249&$-$0.043&$-$0.485&$-$0.44\\ \hline
$\Delta s$&0    &$-$0.064&$-$0.002&$-$0.066&$-$0.10\\ 
\end{tabular}
\end{center}
\end{table}

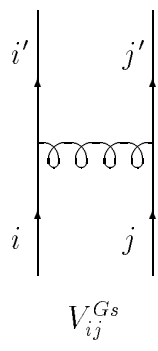
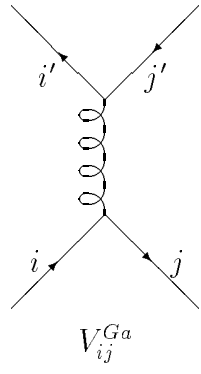
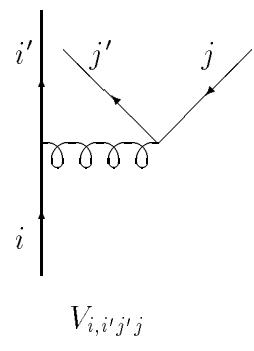
\end{document}